\pdfoutput=1
\documentclass[11pt,twoside,a4paper,cmspaper,final,collab]{cms-tdr}
\def\svnVersion{cd2ebbe}\def\svnDate{2024/11/13}\def\cmsCernNoTag{CERN-EP-2024-059}\def\cmsCernDate{\today}\def\cmsMessage{Published in Physical Review Letters as \href{http://dx.doi.org/10.1103/PhysRevLett.133.201803}{\doi{10.1103/PhysRevLett.133.201803}.}}
\begin{document}\cmsNoteHeader{EXO-21-004}

\ifthenelse{\boolean{cms@external}}{\providecommand{\cmsLeft}{upper\xspace}}{\providecommand{\cmsLeft}{left\xspace}}
\ifthenelse{\boolean{cms@external}}{\providecommand{\cmsRight}{lower\xspace}}{\providecommand{\cmsRight}{right\xspace}}
\DeclareRobustCommand{\PSGh}{{\HepSusyParticle{h}{}{}}\Xspace}

\cmsNoteHeader{EXO-21-004}

\title{Searches for pair-produced multijet resonances using data scouting in proton-proton collisions at \texorpdfstring{$\sqrt{s} = 13\TeV$}{sqrt(s) = 13 TeV}}

\date{\today}

\abstract{
   Searches for pair-produced multijet signatures using data corresponding to an integrated luminosity of 128 \fbinv of proton-proton collisions at $\sqrt{s} = 13 \TeV$ are presented. A data scouting technique is employed to record events with low jet scalar transverse momentum sum values. The electroweak production of particles predicted in $R$-parity violating supersymmetric models is probed for the first time with fully hadronic final states. This is the first search for prompt hadronically decaying mass-degenerate higgsinos, and extends current exclusions on $R$-parity violating top squarks and gluinos.
}

\hypersetup{
pdfauthor={CMS Collaboration},
pdftitle={Searches for pair-produced multijet resonances using data scouting in proton-proton collisions at sqrt(s) = 13 TeV},
pdfsubject={CMS},
pdfkeywords={CMS, multijets, RPV SUSY, Exotica}}

\maketitle 

Discovery of the Higgs boson~\cite{201230,chatrchyan_2013_observation,20121} completed the Standard Model (SM), but many phenomena in particle physics remain unexplained~\cite{Fukuda_1998, PhysRevLett.126.141801, Aaij:2021vac,Barbier:2004ez}. Diverse theories of physics beyond the SM (BSM) have been proposed. Many posit the existence of new resonances that would be produced in high-energy proton-proton ($\Pp\Pp$) collisions and subsequently decay to multi-quark final states~\cite{Barbier:2004ez,Easa:2020mfv}. This Letter presents searches for pair-produced BSM particles promptly decaying to two or three quarks. Pair-produced $R$-parity violating (RPV) supersymmetric (SUSY) particles, which each decay to pairs or triplets of quarks (\PQq), namely higgsinos ($\PSGh\to \PQq \PQq \PQq$), top squarks ($\PSQt\to  \PQq \PQq$), and gluinos ($\PSg\to \PQq \PQq \PQq$) are used as benchmark models~\cite{Barbier:2004ez}. The searches are based on data from $\Pp\Pp$ collisions at the CERN LHC, collected with the CMS detector in 2016--2018, corresponding to an integrated luminosity of 128\fbinv \cite{CMS-LUM-17-003, CMS-PAS-LUM-17-004, CMS-PAS-LUM-18-002}.

\begin{figure*}[htbp]
    \centering
    \includegraphics[width=0.25\textwidth]{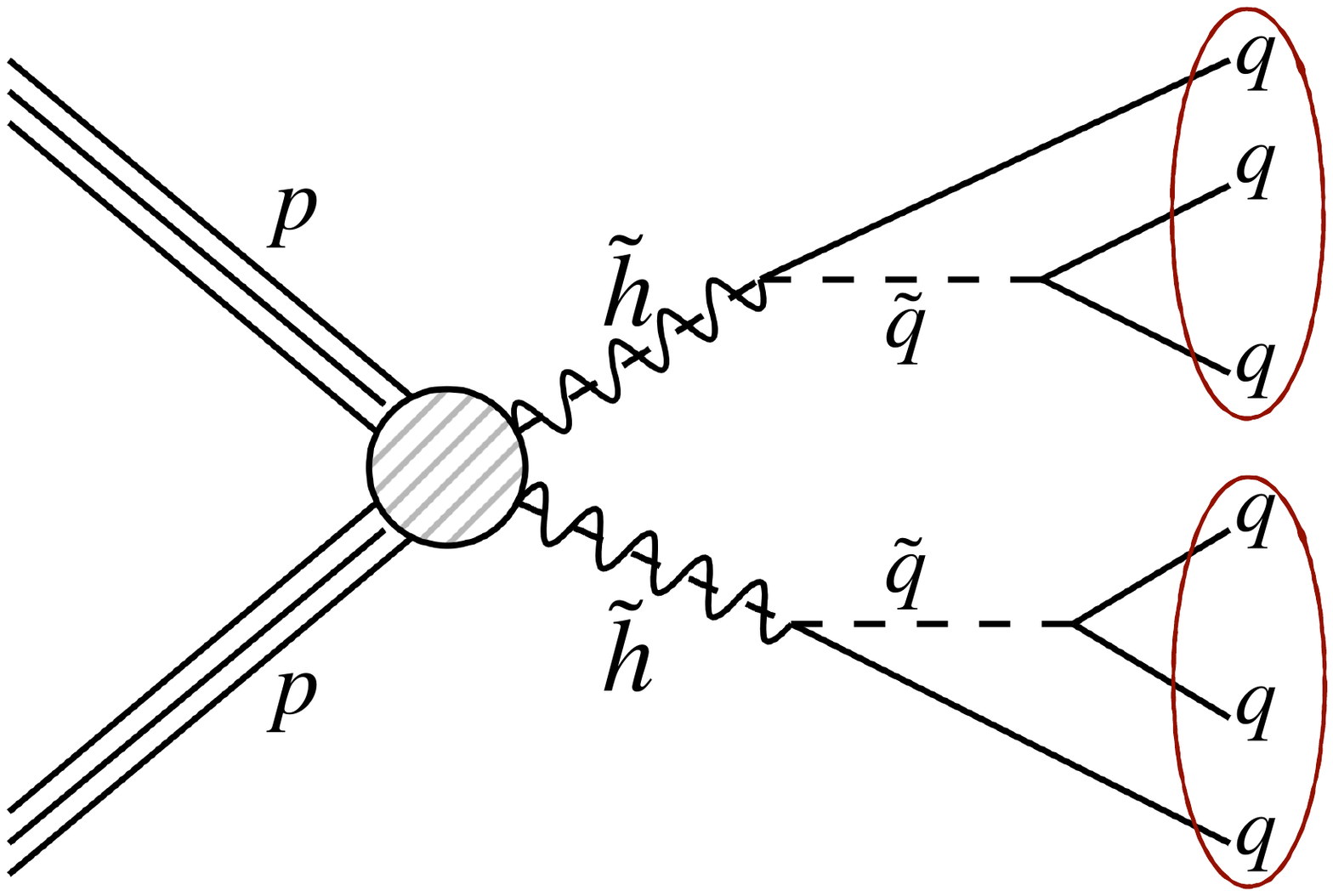} \quad\includegraphics[width=0.25\textwidth]{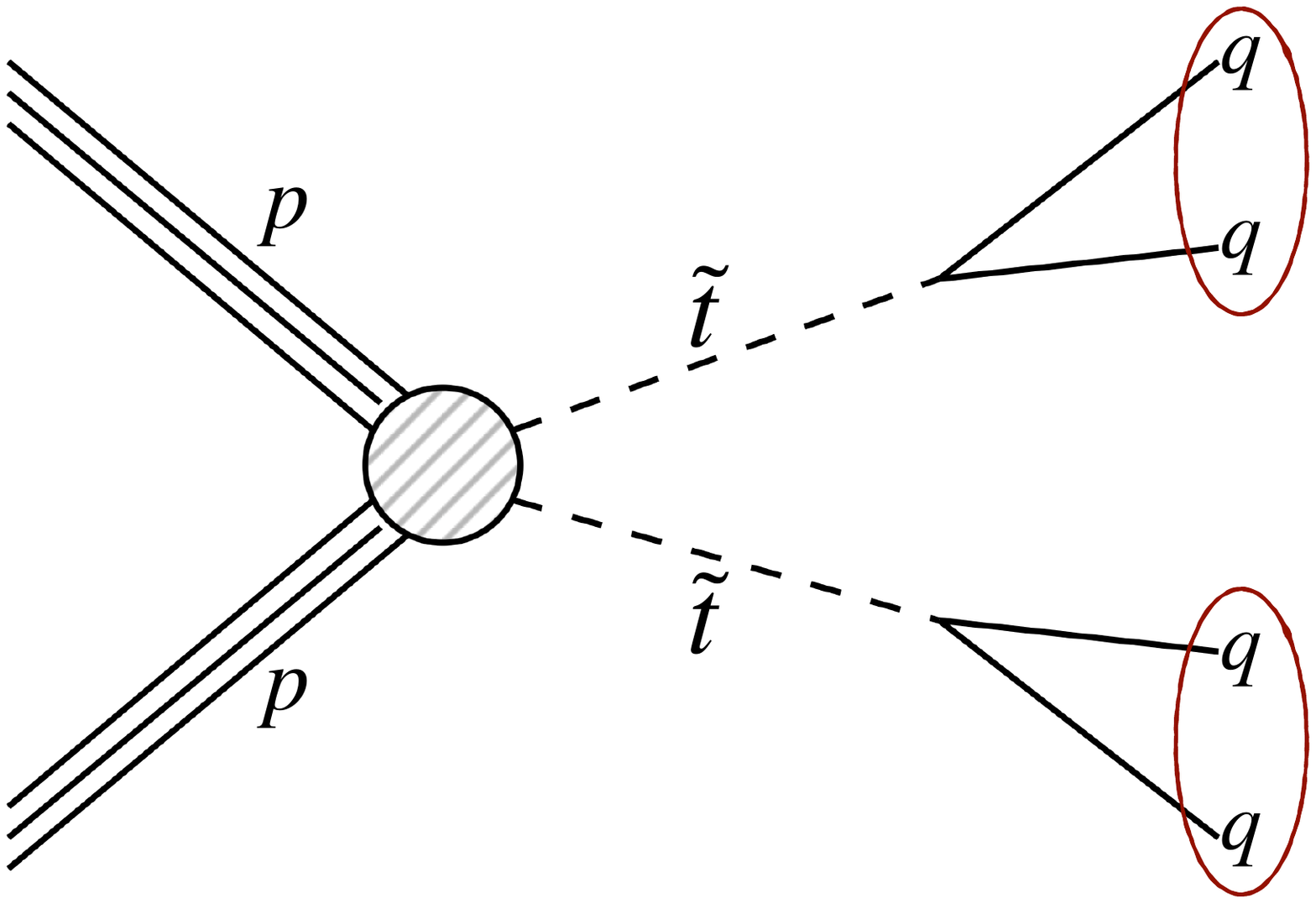}\quad\includegraphics[width=0.25\textwidth]{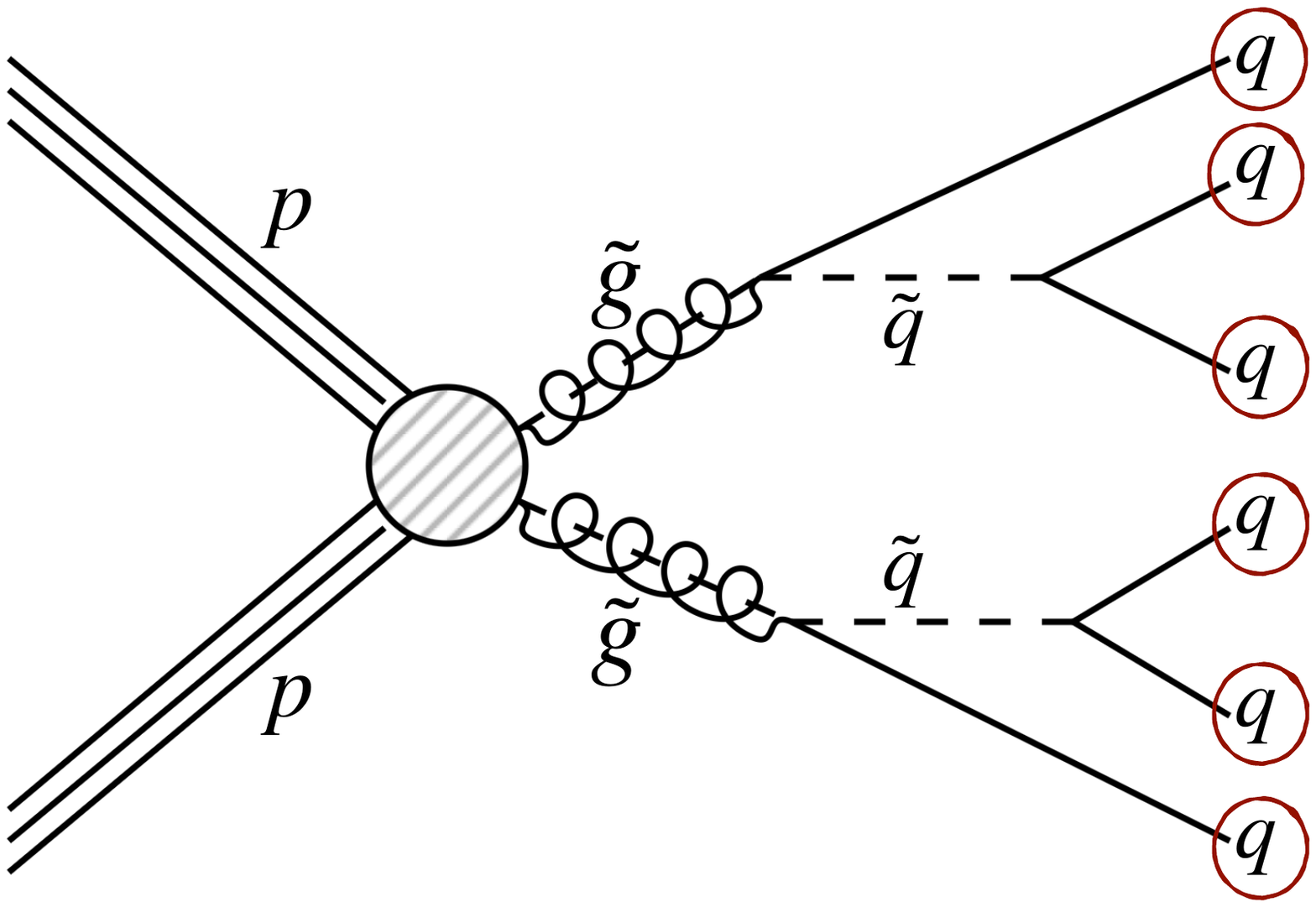}
    \caption{Benchmark RPV SUSY models for merged trijets (left), merged dijets (middle), and resolved trijets (right). The red circles group the final state quarks (or anti-quarks) according to the expected jet clustering of their hadronization products.}
    \label{fig:FDiag}
\end{figure*}

The high center-of-mass energy of 13\TeV of these $\Pp\Pp$ collisions at the LHC can result in low-mass resonances produced with significant Lorentz boosts, such that the hadronization products of the individual final-state quarks overlap in the detector. The resulting signatures are large-radius jets with identifiable substructure. The searches presented in this Letter consider three experimental signatures: pairs of large-radius jets with substructure consistent with three underlying quarks (henceforth called \textit{merged trijets}), pairs of large-radius jets with two-quark substructure (henceforth called \textit{merged dijets}), and pairs of well-resolved triplets of jets (henceforth called \textit{resolved trijets}). Simplified diagrams of the production of each signature are shown in Fig.~\ref{fig:FDiag}. The first two signatures are sensitive to both \PSGh and \PSQt pair production, while the third is sensitive to \PSg pair production.

While the three signatures are analyzed independently, the three analyses use similar techniques for signal selection and share a single method for estimating the dominant background from quantum chromodynamic (QCD) multijet processes. In particular, the analyses use the CMS \textit{scouting data set}~\cite{scouting,CMS:2024zhe}, in which only the event data reconstructed by the high-level trigger (HLT, described below) is retained. The resulting reduced event size allows for higher event rates at lower energy thresholds, thus increasing the sensitivity of searches for lower resonance masses. Jet substructure variables developed explicitly for data scouting are used to reduce the QCD background. The fully hadronic decay of the SM production of top quark-antiquark pairs (\ttbar) is an irreducible background to the paired trijet search. Similarly, events containing hadronically decaying \PZ or \PW bosons are a background to the paired dijet search. These three resonant backgrounds are estimated from simulations.  

Searches for pairs of BSM particles, each decaying to three resolved jets, have been performed by the CDF~\cite{cdfmultijets}, ATLAS~\cite{atlas8, atlas13}, and CMS~\cite{gluino2011,gluino2012,gluino2017} Collaborations. Searches for pairs of particles, which each decay to two quarks, have also been reported by the ATLAS~\cite{atlasstop8,atlasstop13, atlas7} and CMS~\cite{Khachatryan:2014lpa,cmsstop13} Collaborations. While this letter does not present a search for resolved paired dijet resonances, such searches have been performed with the same data set~\cite{Chatrchyan:2013izb, Khachatryan:2014lpa, cmsstop13, R2CMSPDij}. However, no searches have previously been conducted for the production of prompt mass-degenerate RPV higgsinos with a fully hadronic decay or at $\sqrt{s}= 13 \TeV$ for the production of RPV gluinos with masses below 200 \GeV . The searches presented in this Letter are the first to probe these regimes. Tabulated results for this analysis are provided in the HEPData record \cite{hepdata}.

The CMS apparatus~\cite{CMS:2008xjf} is a multipurpose, nearly hermetic detector, designed to trigger on~\cite{CMS:2020cmk,CMS:2016ngn} and identify electrons, muons, photons, and hadrons~\cite{CMS:2020uim,CMS:2018rym,CMS:2014pgm}. A global ``particle-flow'' (PF) algorithm~\cite{CMS:2017yfk} aims to reconstruct all individual particles in an event, combining information provided by the all-silicon inner tracker and by the crystal electromagnetic and brass-scintillator hadron calorimeters, operating inside a 3.8\unit{T} superconducting solenoid, with data from the gas-ionization muon detectors embedded in the flux-return yoke outside the solenoid. The reconstructed particle candidates are used to build \PGt leptons, jets, and missing transverse momentum~\cite{CMS:2018jrd,CMS:2016lmd,CMS:2019ctu}.

While the LHC provides collisions every 25\unit{ns}, the CMS detector and data acquisition systems do not have the bandwidth to record every event. Instead, events of interest are selected using a two-tiered trigger system. The first level (L1), composed of custom hardware processors, uses information from the calorimeters and muon detectors to select events at a rate of around 100\unit{kHz} within a fixed latency of about 4\mus~\cite{CMS:2020cmk}. The second level, known as the HLT, consists of a farm of processors running a version of the PF algorithm optimized for fast processing, and reduces the event rate to around 1\unit{kHz} before data storage~\cite{CMS:2016ngn}. If the partial reconstruction by the HLT indicates that the event has passed some specific requirements, such as the scalar sum of the transverse momenta (\pt) of jets in the event (\HT) being above some threshold, the event is fully reconstructed and stored for offline analysis.

To handle the high multijet production rates of the LHC, the threshold for standard trigger selection based purely on \HT was set to 800\GeV in 2016 and 1.05 \TeV in 2017--2018. For the range of new particle masses considered in this Letter, the resonances rarely decay to jets with sufficient \pt for the events to meet these thresholds. To allow analyses targeting low-mass final states to overcome these limitations, CMS implemented data scouting, in which only the objects reconstructed at the HLT are saved for further analysis~\cite{CMS:2024zhe}. In particular, the analyses use a data set of scouting events with $\HT > 410 \GeV$~\cite{CMS:2017yfk, scouting}. Data scouting allows CMS to collect data at a rate much higher than that possible with standard triggers. During one of the highest instantaneous luminosity data collection periods, the scouting triggers used in these searches were able to collect data at 1.8\unit{kHz}, while the rate for comparable standard triggers was 400\unit{Hz}. The data size of scouting events is approximately 60 times smaller than that of standard events.

Pair production of RPV SUSY particles is used as the benchmark model for these analyses, where gluinos and higgsinos decay to the three-jet final state and top squarks decay to the dijet final state. Signal events are generated using \MGvATNLO 2.6.5~\cite{madgraph} with up to one additional jet using the MLM matching procedure~\cite{mlmmatch}. The masses of virtual squarks in gluino decays are set to sufficiently high values that the gluinos decay without an internal resonance. The natural width of the RPV higgsinos,  RPV gluinos, and RPV top squarks is assumed to be much smaller than the experimental resolution. QCD and~\ttbar events, as well as events with \PW and \PZ bosons produced in association with extra jets (\PW/\PZ+ jets) are simulated, with \MGvATNLO 2.6.5. QCD and \PW/\PZ+ jets events are simulated at leading order with up to four extra partons in the matrix element calculations, while~\ttbar events are simulated at next-to-leading order. 

All samples were interfaced with \PYTHIA 8.212~\cite{pythia} with CP5 as the underlying event tune~\cite{cp5} using the parton distribution function set NNPDF3.1NNLO~\cite{Ball:2014uwa}. The interaction of particles with the detector is simulated using the \GEANTfour toolkit~\cite{Agostinelli:2002hh}. The effects of particles from additional $\Pp\Pp$ interactions in the same or adjacent bunch crossings (pileup) are incorporated by overlaying minimum bias events, simulated with \PYTHIA, on the hard scattering interactions, with the multiplicity distribution of these additional interactions matching that observed in data. The QCD simulation is only used to optimize the event selection procedure for each analysis. The QCD background for each search is estimated using Gaussian process (GP) regression as described below\cite{Gandrakota:2022wyl}.  

All the PF candidates in an event are clustered into jets using \FASTJET~\cite{Cacciari_2012,Cacciari:2005hq} with the anti-\kt algorithm~\cite{Cacciari_2008} and distance parameters of 0.4 and 0.8 (AK4 and AK8 jets) for the resolved and merged categories, respectively. The jet energies are corrected to compensate for the combined response functions of the CMS calorimeters and the effects of pileup~\cite{CMS:2016lmd,CMS:2020ebo}. Jets in data are further corrected to account for the residual difference between simulation and data. The corrections are derived from data that has undergone full offline reconstruction, with small residual corrections appropriate for the HLT reconstruction used in scouting dataset. The tight jet identification criteria~\cite{jetid} are applied to remove poorly reconstructed jets and calorimeter noise misidentified as jets.

The trimming algorithm described in Ref.~\cite{Krohn:2009th} is applied to the AK8 jets. This algorithm clusters the constituents of the jet into smaller jets with radius of 0.2 using the \kt algorithm. Clusters with \pt smaller than $3\%$ of the original jet \pt are removed. This trimming procedure reduces contributions from soft QCD radiation, pileup, and particles from the underlying event to the jet mass $m$. After trimming $m$ is further corrected so that the top quark mass distribution observed in data matches that from \ttbar simulation.

In the merged analyses, jet substructure variables~\cite{Thaler:2010tr} are used to separate signal jets (AK8 jets containing the hadronization of multiple quarks) from the QCD background. In particular, the variables $\tau_{32}$~\cite{Thaler:2010tr} and $N^1_2$~\cite{Moult:2016cvt} are used to identify resonances decaying to three or two quarks with merged final states, respectively. These variables have lower values when evaluated on jets resulting from overlapping quark decays, and higher values for jets initiated by a single quark or gluon. These substructure variables are correlated with the kinematic properties of the jet, such that imposing a selection based on them can preferentially enhance regions of the background mass distributions. To avoid these correlations, the ``designed decorrelated tagger'' (DDT)~\cite{Dolen:2016kst} procedure is used to reduce the dependence of the variables on the \pt and $\rho=\ln{(m^2/\pt^2)}$ of the jets. The decorrelation is achieved by varying the substructure variables in each \pt and $\rho$ bin such that 5\% of simulated QCD events are allowed in each bin. The resulting variables are called $N^1_{2,\text{DDT}}$ and $\tau_{32,\text{DDT}}$. This methodology has been used in several searches performed by the CMS Collaboration~\cite{CMS:2017dcz, CMS:2019emo, CMS:2019xai}. Both the leading and sub-leading jets are required to have $\tau_{32,\text{DDT}}<0$ or $N^1_{2,\text{DDT}}<0$, for selecting pair-produced resonances decaying to three or two quarks, respectively.

For the merged resonance searches, only the leading two AK8 jets in the events are considered. To ensure that the analyses only consider events for which the trigger is fully efficient, each jet is required to have $\pt > 300 \GeV$ and $\abs{\eta}<2.4$. The mass asymmetry between the leading jet ($j_1$) and sub-leading jet ($j_2$) sorted by jet \pt, $A_m \equiv \abs{m(j_1)-m(j_2)}/(m(j_1)+m(j_2))$ is required to be less than 0.15.

For the resolved analysis, a neural network (NN) based quark-gluon discriminator (QGD) is used to distinguish gluon-initiated AK4 jets from the quark-initiated AK4 jets expected for trijet resonances. The network's architecture is based on the "Particle-Flow network" as described in Ref. ~\cite{Komiske:2018cqr}. The NN inputs are the normalized four-momenta information ($\pt,\eta,\phi,m$) along with the particle type of each jet constituent. Each constituent's \pt is divided by the \pt of the entire jet. The NN is trained on an equal number of quark and gluon jets, sampled from the QCD simulation using truth-level information, for a total of 900,000 jets. The NN outputs a QGD score between 0 and 1, with 1 corresponding to more quark-like jets. The analysis uses loose, medium, and tight QGD selections corresponding to the quark jet acceptance and gluon jet rejection rates of 98\% and 31\%,  83\% and 70\%, and 61\% and 87\%, respectively. The tight and loose selections are optimized for best quark jet acceptance and gluon jet rejections efficiencies, respectively.

The analyses must contend with immense background from QCD multijet processes. In the case of the merged signatures, jet substructure is used to suppress this background. The sensitivity of the resolved trijet search is improved by using the QGD with the jet-ensemble technique, which examines all possible combinations of three jets as described in Ref.~\cite{gluino2017}. The remaining QCD background in the analyses is estimated using GP regression~\cite{gpml}, a technique for fitting data without assuming a prior for functional form, utilized for these kinds of searches for the first time. Backgrounds from \ttbar are estimated using Monte Carlo (MC) simulation. 

Events in the resolved trijets search are divided into three regions, based on the targeted gluino mass. The three mass ranges are given in Table \ref{table:optpt}. For all three search regions, in the event reconstruction all AK4 jets that pass the thresholds of $\pt>30\GeV$ and $\abs{\eta}<2.4$ are initially considered. The events are required to have a total $\HT>600\GeV$, to ensure a fully efficient trigger. Jets that fail the loose quark threshold of the QGD are rejected, and every event is required to have at least six jets passing this selection criteria. In events with more than six jets, it was found that restricting the set of considered jets to the six with the highest \pt maximized the sensitivity of the analysis. These six jets are grouped into combinations of three, making twenty triplets, and ten triplet pairs.

The final selection for each of the three gluino mass regions is based on six variables, as described in Ref.~\cite{gluino2017} . The lowest \pt of the six jets must be above a given threshold.  The event-level Dalitz variable $D^2_{[(6,3)+(3,2)]}$, which measures the geometric spread in the six-jet topology, cannot exceed a given threshold. The $A_m$ value between the constituent triplet objects cannot exceed a predefined threshold. A variable $\Delta \equiv p_{T_{jjj}} - m_{jjj}$, is defined for triplets from the remaining pairs.  Trijets from QCD multijets have a low value of $\Delta$, as $p_{T_{jjj}}$ and $m_{jjj}$ scale with each other.   A triplet-level requirement on the Dalitz variable $D^2_{[3,2]}$ is imposed to enforce a symmetry of decay in the triplets. Finally, all jets within the triplets passing the above criteria are required to pass the medium QGD selection and at least one jet must pass the tight QGD selection. The selection variables were optimized individually for each of the three gluino mass ranges. The values of the relevant thresholds are shown in Table \ref{table:optpt}. Events must have at least one such triplet, and if multiple jet triplets in an event fulfill these requirements, all such trijet combinations are considered in the final distribution of reconstructed gluino masses.

\begin{table*}[htb!]
\centering
\topcaption{  Selection requirements for the resolved trijet resonance search listed for the three regions of three-jet resonance mass.}
\begin{scotch}{cclllllll }
    Region & Trijet mass  & \multicolumn{1}{c}{6th jet \pt} & \multicolumn{1}{c}{$D^2_{[(6,3)+(3,2)]}$} & \multicolumn{1}{c}{$A_m$} & \multicolumn{1}{c}{$\Delta$} & \multicolumn{1}{c}{$D^2_{[3,2]}$} \\ 
      & (\GeVns)   & \multicolumn{1}{c}{(\GeVns)} &  &  & (\GeVns) &  \\\hline
    1 & 200--500 & $>$40& $<$1.25 & $<$0.25 & $>$250& $<$0.05\\
    2 & 500--900 & $>$50& $<$1.00 & $<$0.175 & $>$180& $<$0.175\\
    3 & 900--2000 & $>$100& $<$0.9 & $<$0.15 & $>$150& $<$0.2 \\
  \end{scotch}
  \label{table:optpt}
\end{table*}

The GP regression technique is used to estimate the dominant background from QCD~\cite{Gandrakota:2022wyl}. Unlike the rigid constraints imposed by conventional curve fitting approaches, GP is a flexible alternative that does not specify a parametric functional form, instead providing a probability distribution over possible functions that fit the data~\cite{gpml}. The GP regression is defined using a kernel that directly encodes an understanding of the underlying physics, manifesting as controlled covariance among the bin counts~\cite{frate2017modeling}. In these analyses the radial bias function (RBF) kernel is used to model the QCD multijet invariant mass distributions, as described in Ref.~\cite{Gandrakota:2022wyl}. This kernel describes the correlation between bins in the mass spectra by assigning an exponential weight that falls with the square of their separation. Use of the GP regression technique was previously observed to give accurately modeled multijet mass distributions~\cite{frate2017modeling}. The signal shapes are modeled by Gaussian distributions. A bias test was performed at various masses by injecting signal shapes to validate the fitting procedure, and yielded no indication of significant bias in the signal extraction process. The background contributions from processes with SM multijet resonances (\ttbar, \PW/\PZ+ jets) are taken from simulation, with their cross sections constrained by the GP regression in data. All other SM background contributions are found to be negligible.

\begin{figure}[htb!]
    \centering
    \includegraphics[width=0.32\textwidth]{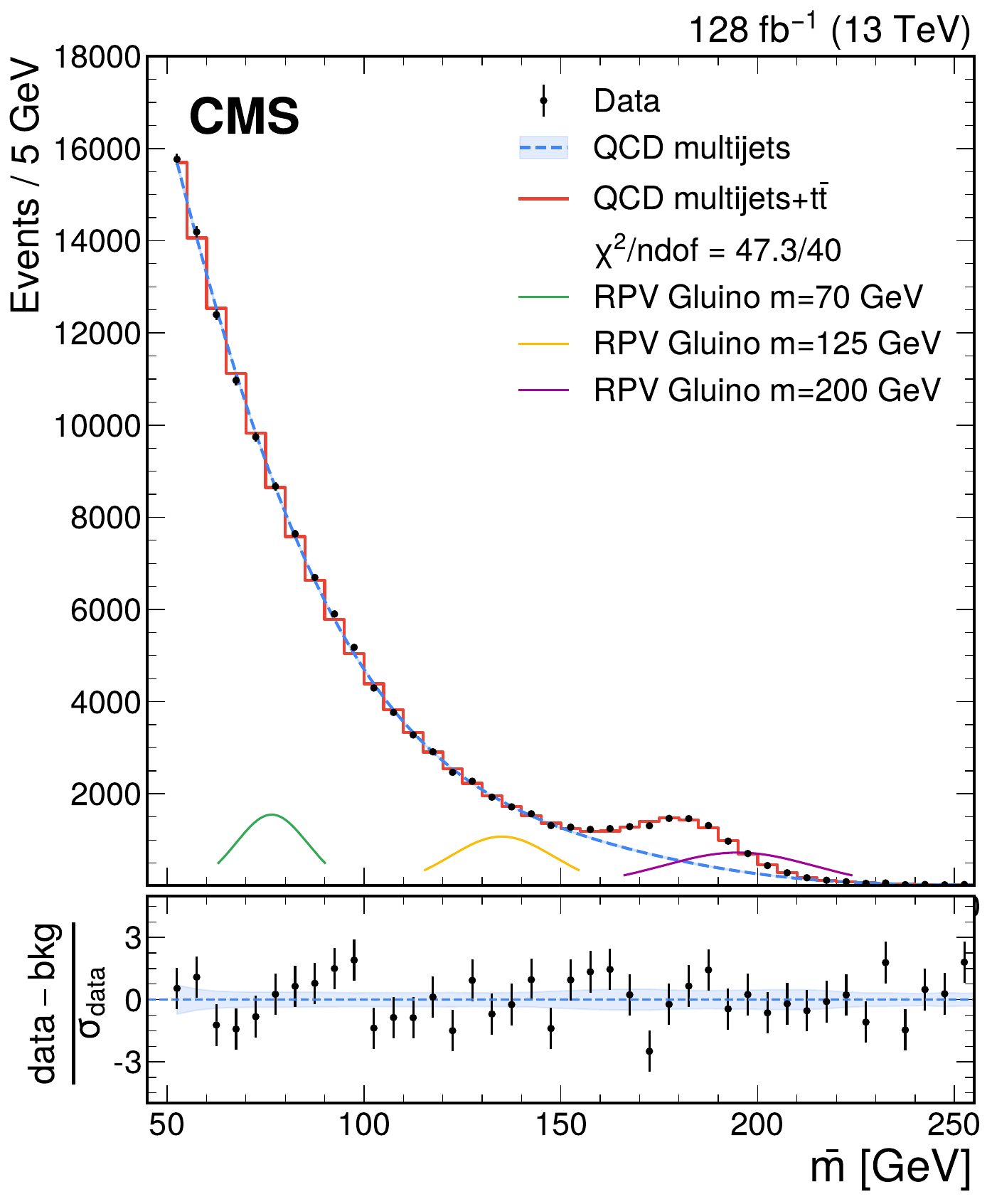}
    \includegraphics[width=0.32\textwidth]{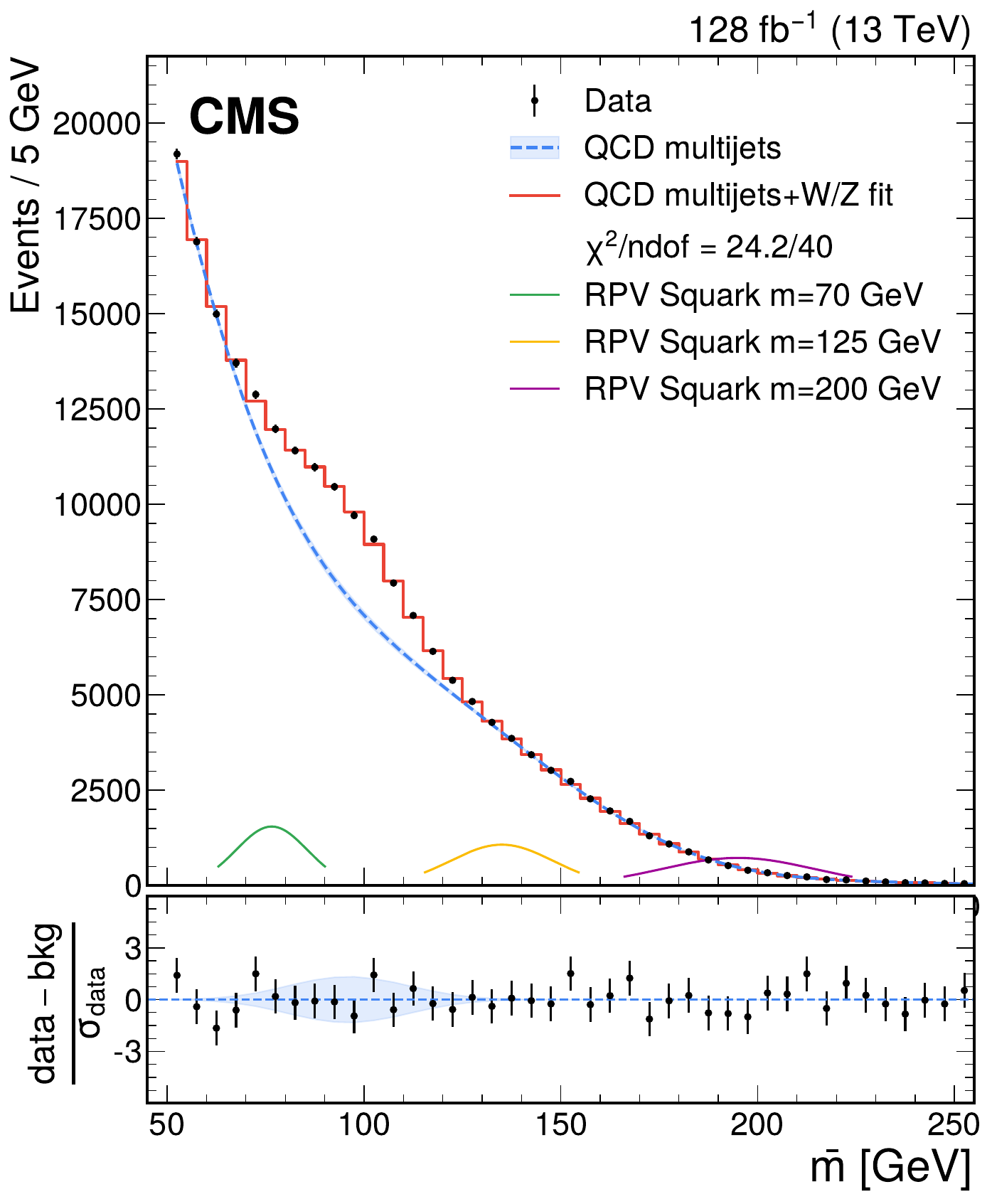}
    \caption{The distribution of average jet mass in the data (points), for the search for pair-produced merged resonances decaying to trijets (\cmsLeft) and dijets (\cmsRight), compared to a background-only prediction from GP regression (blue), and the full background fit including simulations (red) of SM resonances from \ttbar (\cmsLeft) and \PW/\PZ+ jets (\cmsRight). Also shown are the expected shapes of signals from $R$-parity violating gluinos or squarks with resonance masses 70 (green), 125 (yellow) and 200\GeV(purple), with arbitrary normalizations. The bottom panel shows the difference between the data and the final background estimate, divided by the statistical uncertainty of the data in each bin, with fit uncertainty shown in blue.}
    \label{fig:background_prediction}
\end{figure}

\begin{figure*}
    \centering
    \includegraphics[width=0.3\textwidth]{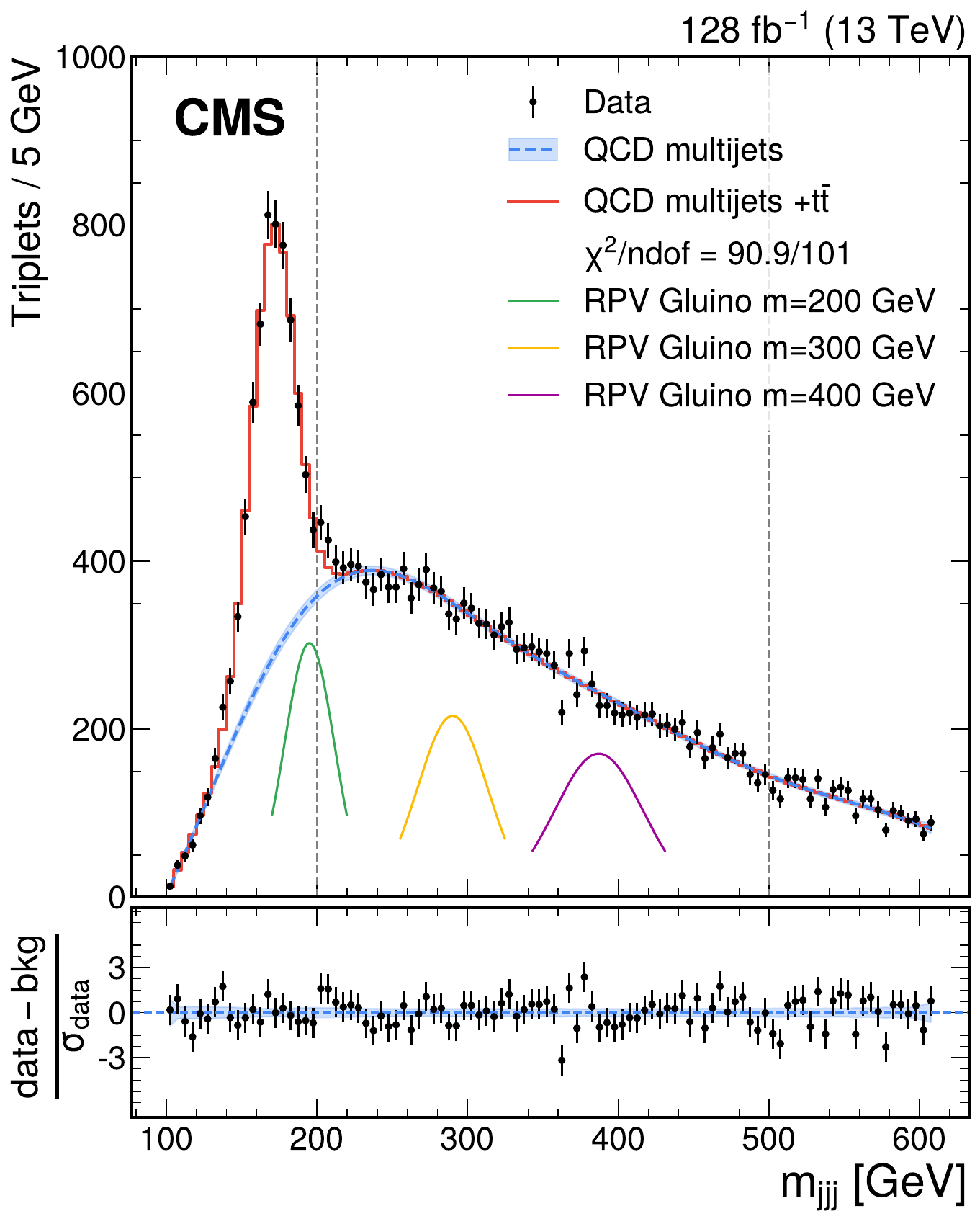}
    \includegraphics[width=0.3\textwidth]{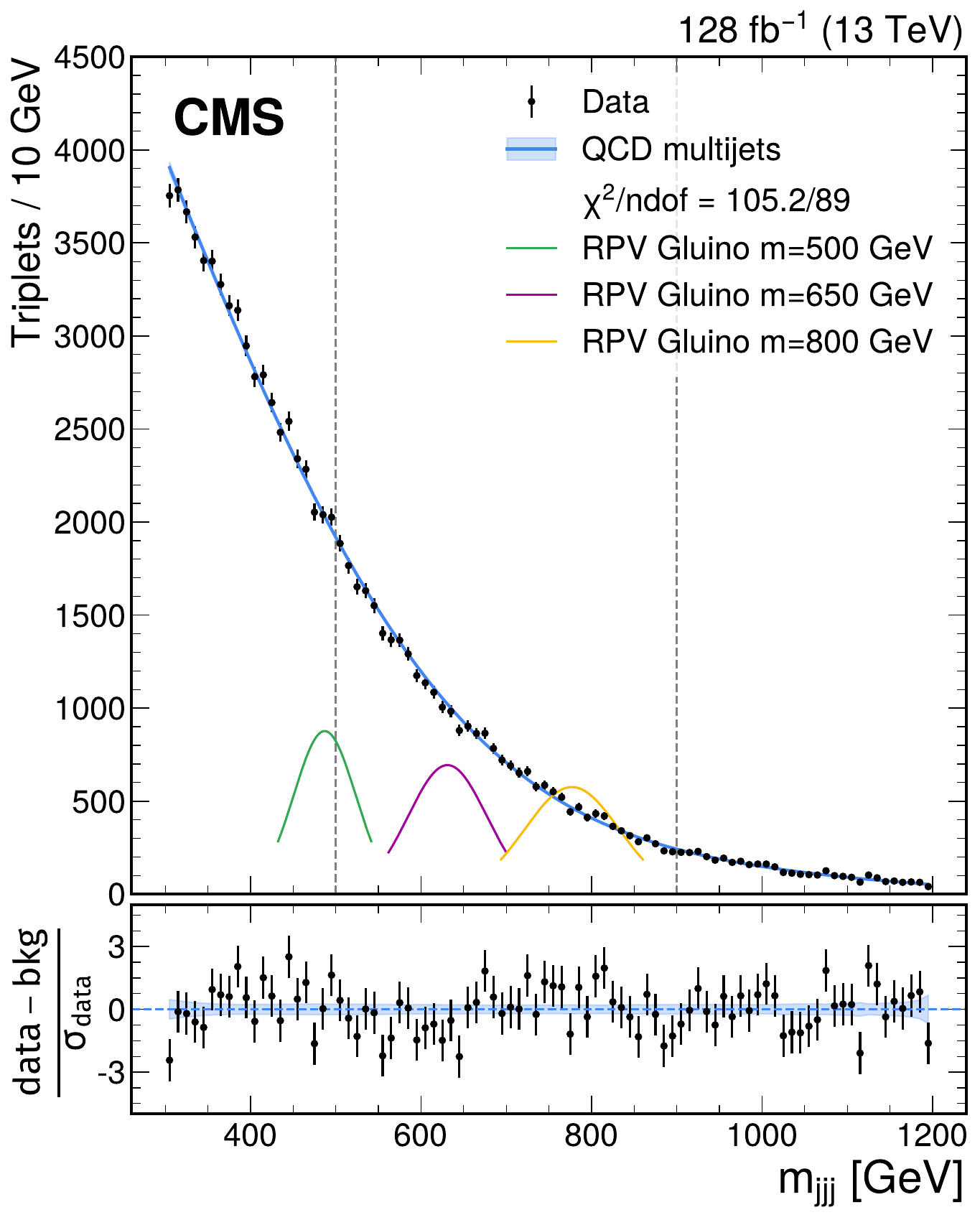}
    \includegraphics[width=0.3\textwidth]{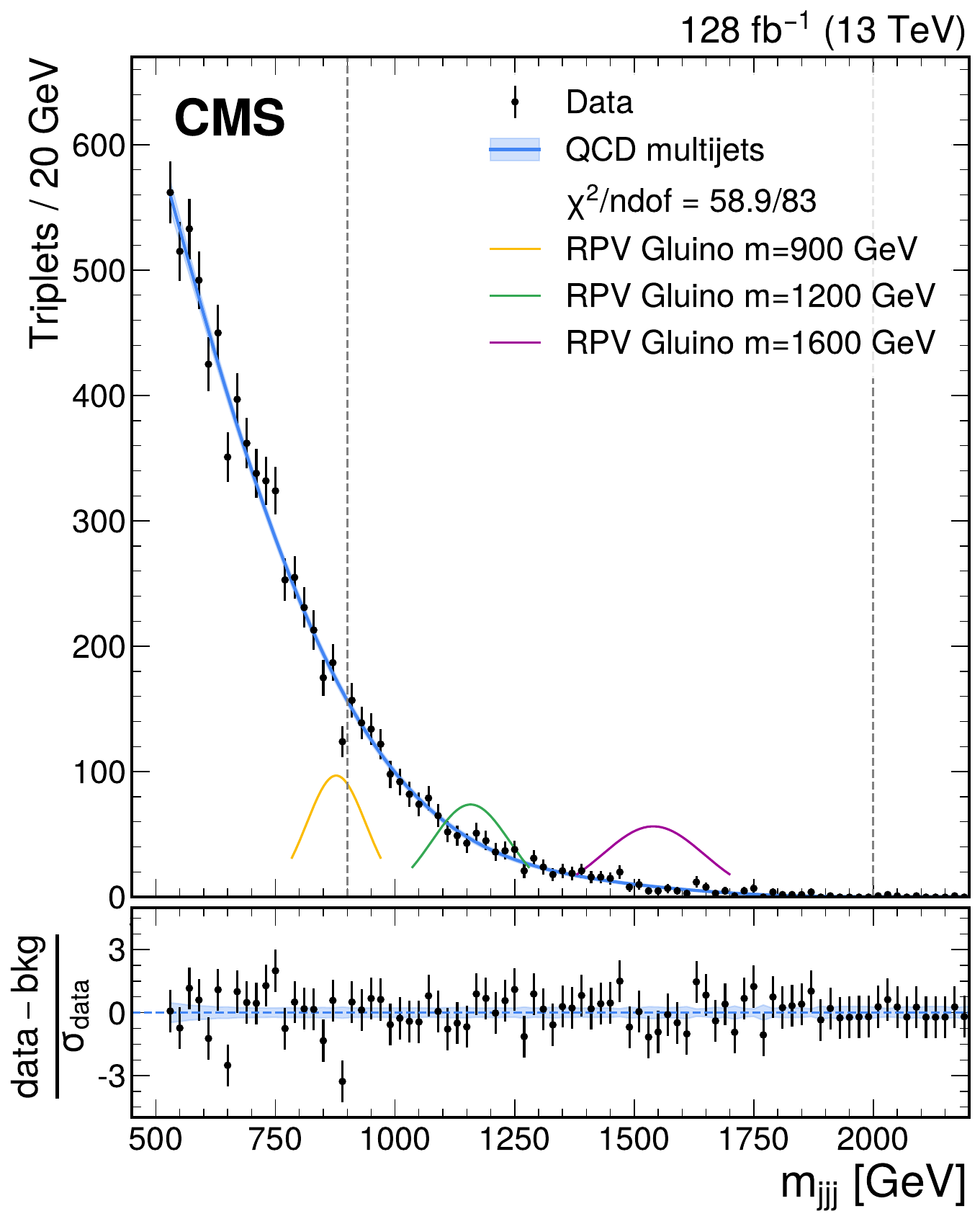}
    \caption{The distribution of trijet mass in the data (points), for the three search regions (\cmsLeft, middle and \cmsRight) for pair-produced resolved resonances decaying to trijets, compared to a background-only fit from GP regression (blue), and (\cmsLeft) the full background fit including simulation of a SM resonance from \ttbar production (red). Also shown are the expected shapes of signals from $R$-parity violating gluinos at various masses (green, yellow and purple), with arbitrary normalizations. The search range is indicated by the vertical dashed lines. The bottom panel shows the difference in each bin between the data and the final background estimate, divided by the statistical uncertainty of the data, with fit uncertainty shown in blue.}
    \label{fig:resolved_bkg}
\end{figure*}

For the merged trijet and merged dijet searches, this background estimation procedure is applied to the average jet mass ($\overline{m}=(m_1+m_2)/2$) distribution of events passing the selection criteria for pair-produced merged resonances decaying to three or two quarks, respectively, as illustrated in Fig.~\ref{fig:background_prediction}. Similarly, in the resolved trijet search, this background estimation procedure is applied to the individual trijet invariant mass ($m_{jjj}$) distributions for jet triplets passing the selection criteria, as illustrated in Fig.~\ref{fig:resolved_bkg}. The GP regression and added \ttbar and \PW/\PZ simulations describe the data well, with $p$-values using a $\chi^2$ metric of 0.75, 0.12, and 0.97 for the three regions of the resolved trijet search, and 0.21 and 0.96 for the merged trijet and merged dijet searches, respectively. 

The strength of the selection criteria is indicated by the clear observation of the \ttbar and \PW/\PZ signals in the merged trijet and dijet resonance search (Fig.~\ref{fig:background_prediction}) and of the \ttbar signal in the resolved trijet search (Fig.~\ref{fig:resolved_bkg}). The rates of \ttbar and \PW/\PZ production observed in the trijet and dijet searches are within 7\% and 6\% of SM expectations, respectively. 
 
The widths of the Gaussian distributions used to model the signal are extracted from the  mass spectra of the simulated higgsinos, top squarks, and gluinos. We observed an effectively linear relation between the width of the Gaussian and the resonance mass for our search ranges. The widths are found to be 9.1\% and 7.5\% of resonance mass in the merged and resolved searches, respectively. The search is performed by fitting these signal templates and the backgrounds simultaneously to the data using a Bayesian approach described in detail below.  No significant deviation from the SM is found; the largest excess occurs at a reconstructed trijet mass of approximately 720 \GeV as shown in Fig.~\ref{fig:resolved_bkg} (middle) and has a local significance of 2.6 standard deviations. This corresponds to a gluino mass of 770 \GeV.

This analysis uses a Bayesian procedure with a uniform prior signal strength to evaluate significance and determine the 95\% confidence level (\CL) upper limit~\cite{CowanPDGStat,CMS-NOTE-2011-005}. The Markov chain MC (MCMC) method is used to marginalize the nuisance parameters~\cite{Gandrakota:2022wyl,emcee}. 
The MCMC sampling procedure is applied to 1,000 pseudo data sets. The posterior distributions from these are used to obtain the estimated limits.

The uncertainty in the selection efficiency of the substructure variables is measured using semi-leptonic decays of \ttbar. This is the dominant systematic uncertainty. The uncertainty in the integrated luminosity measurement also contributes to the uncertainty in the signal rate. The contribution from the choice of parton distribution functions for the signal simulations was found to have a negligible effect. All the signal rate uncertainties are modeled with log-normal probability density functions (PDFs). The uncertainties arising from the jet energy corrections (JEC) and the uncertainty in the jet energy resolution (JER) both affect the shape of the signal. These are modeled as nuisance parameters affecting the mean and width of the signal, respectively. The nuisance parameters corresponding to the signal and background shape uncertainties are modeled with Gaussian PDFs. The shape and rate uncertainties that affect the signal also affect the \ttbar and \PW/\PZ backgrounds. The standard deviations for all nuisance parameters are shown in Table \ref{sys_table}.

\begin{table}[htb!]
\centering
\topcaption{Summary of the systematic uncertainties in the signal yield and shape, for the searches for merged three and two quark, and resolved trijet resonances.}
\label{sys_table}
\begin{scotch}{lcccc}
Source of uncertainty              & Effect          &Merged&Merged& Resolved\\
             &           &trijets&dijets&trijets\\ \hline
Trigger efficiency                & Yield           & 2.5\%          & 2.5\%          & 2.5\%            \\
Acceptance                        & Yield           & 5\%            & 5\%            & 5 \%             \\
JEC             & Mean  & 3.5\%          & 3.5\%          & 3.5\%            \\
JER             & Width & 12\%           & 12\%           & 12\%             \\
QGD efficiency                    & Yield           & \NA              & \NA              & 5\%              \\
$\tau_{32,\text{DDT}}$ efficiency & Yield           & 4\%            & \NA              & \NA                \\
$N^1_{2,\text{DDT}}$ efficiency   & Yield           & \NA              & 20\%           & \NA       
\end{scotch}
\end{table}

\begin{figure*}[htbp!]
    \centering
    \includegraphics[width=0.32\textwidth]{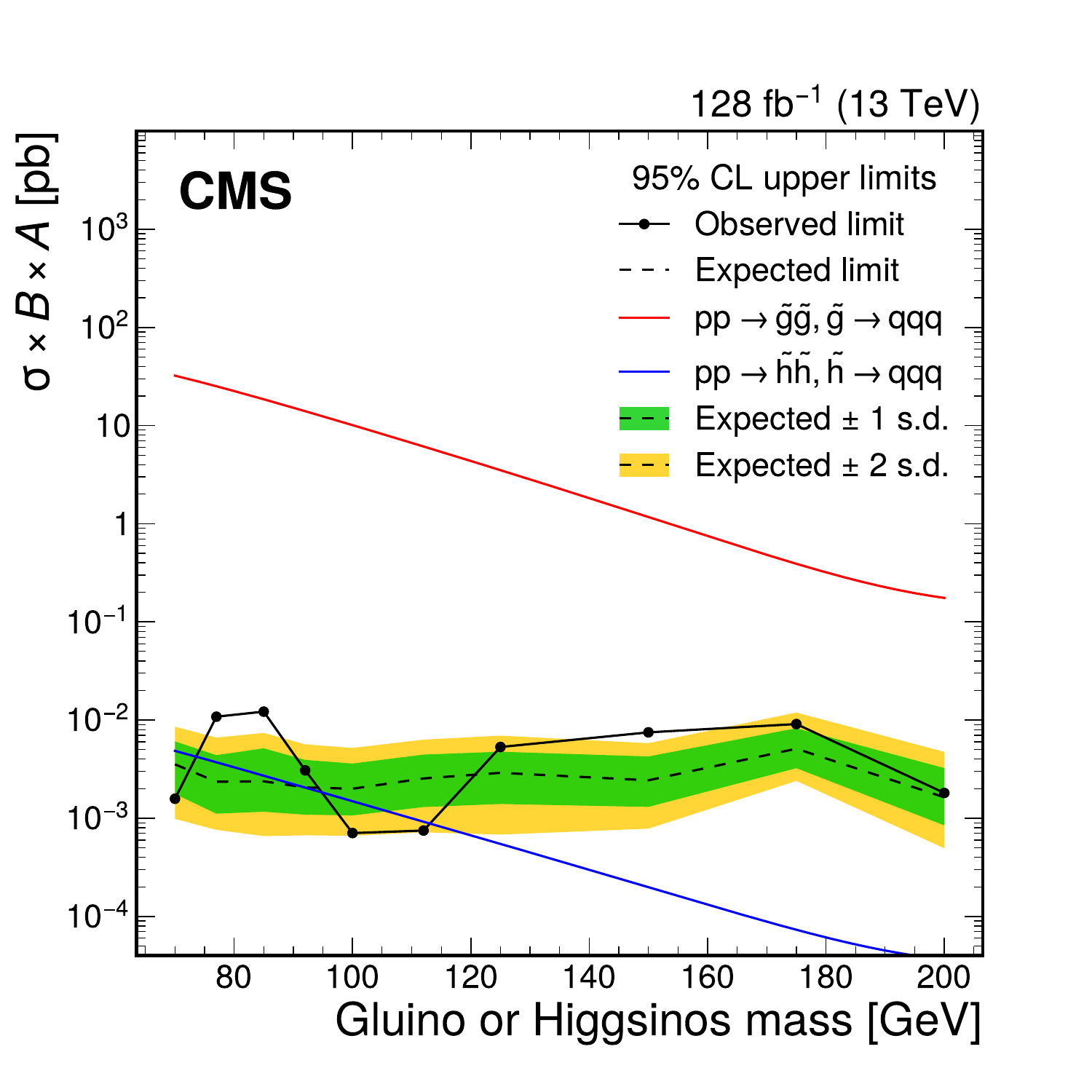}
    \includegraphics[width=0.32\textwidth]{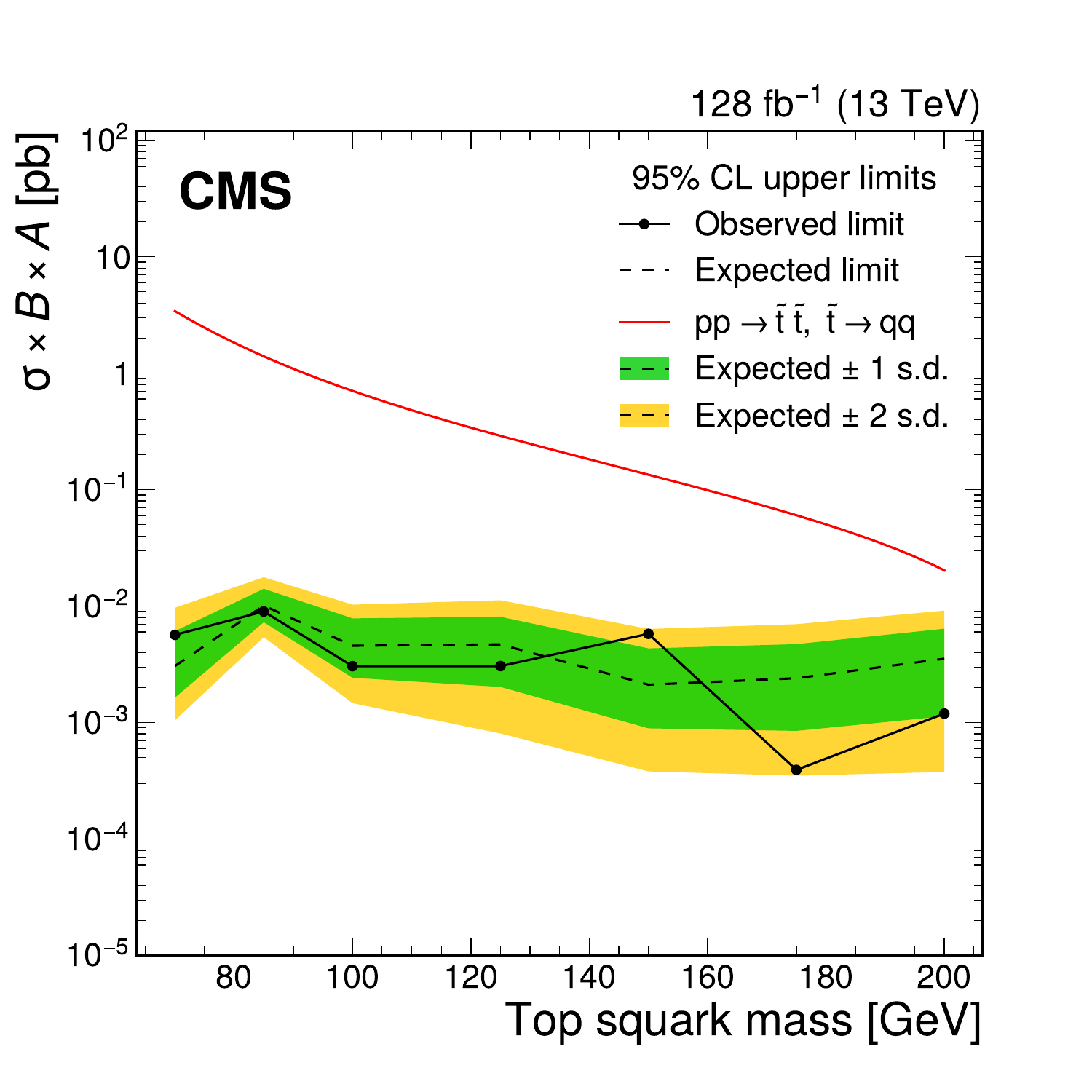}
    \includegraphics[width=0.32\textwidth]{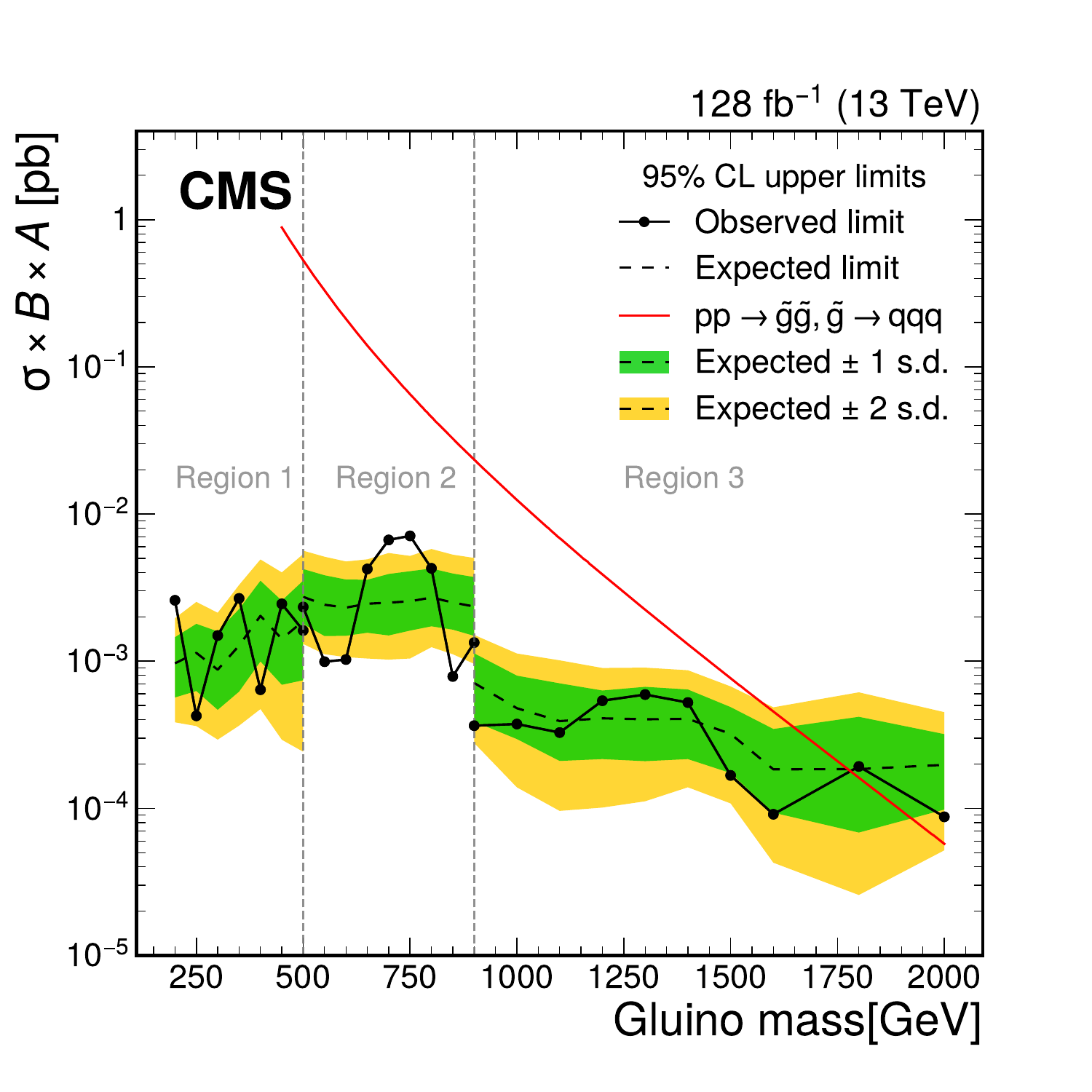}
    \caption{Observed limits (points) and expected limits (dashed curves) on the product of the production cross section ($\sigma$), branching fraction (B), and acceptance (A) for pair produced merged trijets (\cmsLeft), merged dijets (middle), and pair produced resolved trijets (\cmsRight). The vertical lines on the resolved three quark resonance limits (\cmsRight) indicate the different search regions. The limits are compared to theoretical predictions for the pair production of $R$-parity violating gluinos (red in \cmsLeft and \cmsRight), mass degenerate higgsinos (blue in \cmsLeft), and top squarks (red in middle).}
    \label{fig:limits}
\end{figure*}

Upper limits on the product of production cross section, branching fraction, and acceptance for each signature are shown in Fig.~\ref{fig:limits}. The results were compared to asymptotic \CLs~\cite{Cowan:2010js,CLS1,CLS2} limits for a subset of signal masses and were found to be in close agreement. The acceptance after all selection criteria varies between $10^{-5}$ and $6\times10^{-5}$, $4\times10^{-5}$ and $1.7\times10^{-4}$, and $2\times10^{-3}$ and $6\times10^{-2}$ for the merged three quark, merged two quark, and resolved three quark resonance searches respectively.  The substructure requirements in merged resonance searches, and the $\Delta$ selection in the resolved trijet search are the principal causes of the diminished acceptance rates. The upper limits are compared to predictions for RPV gluinos, RPV top squarks, and mass-degenerate RPV higgsinos~\cite{Borschensky:2014cia}, assuming a 100\% branching fraction to quarks. The merged trijet resonance search sets the first limits on mass degenerate RPV higgsinos, excluding masses between 70 and 75\GeV and between 95 and 112\GeV at 95\% \CL. The merged dijet resonance search excludes RPV top squarks with masses between 70 and 200 \GeV at 95\% \CL, extending the exclusion to lower mass and yielding higher sensitivity than previous results~\cite{cmsstop13}. Finally, the trijet searches exclude RPV gluinos with masses between 70 GeV and 1.7\TeV, and supersede earlier results~\cite{gluino2017}.

In summary, a search has been performed for pair-produced multijet resonances using data scouting, where the jets in the final state can be either merged or resolved individually. This search accurately reconstructs the hadronic decays of the top quark and \PW, \PZ bosons, in agreement with  SM expectations. New, additional resonances were not observed, with the largest excess consistent with a three jet resonance mass of 770 \GeV, but with a local significance of only 2.6 standard deviations. This search extends the previous limits on $R$-parity violating models of top squarks and gluinos, and sets the first exclusion limits on $R$-parity violating mass-degenerate, prompt, and hadronically decaying higgsinos.

\clearpage

\begin{acknowledgments}
We congratulate our colleagues in the CERN accelerator departments for the excellent performance of the LHC and thank the technical and administrative staffs at CERN and at other CMS institutes for their contributions to the success of the CMS effort. In addition, we gratefully acknowledge the computing centers and personnel of the Worldwide LHC Computing Grid and other centers for delivering so effectively the computing infrastructure essential to our analyses. Finally, we acknowledge the enduring support for the construction and operation of the LHC, the CMS detector, and the supporting computing infrastructure provided by the following funding agencies: SC (Armenia), BMBWF and FWF (Austria); FNRS and FWO (Belgium); CNPq, CAPES, FAPERJ, FAPERGS, and FAPESP (Brazil); MES and BNSF (Bulgaria); CERN; CAS, MoST, and NSFC (China); MINCIENCIAS (Colombia); MSES and CSF (Croatia); RIF (Cyprus); SENESCYT (Ecuador); ERC PRG, RVTT3 and MoER TK202 (Estonia); Academy of Finland, MEC, and HIP (Finland); CEA and CNRS/IN2P3 (France); SRNSF (Georgia); BMBF, DFG, and HGF (Germany); GSRI (Greece); NKFIH (Hungary); DAE and DST (India); IPM (Iran); SFI (Ireland); INFN (Italy); MSIP and NRF (Republic of Korea); MES (Latvia); LMTLT (Lithuania); MOE and UM (Malaysia); BUAP, CINVESTAV, CONACYT, LNS, SEP, and UASLP-FAI (Mexico); MOS (Montenegro); MBIE (New Zealand); PAEC (Pakistan); MES and NSC (Poland); FCT (Portugal); MESTD (Serbia); MCIN/AEI and PCTI (Spain); MOSTR (Sri Lanka); Swiss Funding Agencies (Switzerland); MST (Taipei); MHESI and NSTDA (Thailand); TUBITAK and TENMAK (Turkey); NASU (Ukraine); STFC (United Kingdom); DOE and NSF (USA).
\end{acknowledgments}

\ifthenelse{\boolean{cms@external}}{\clearpage}{}

\bibliography{auto_generated} \cleardoublepage \appendix\section{The CMS Collaboration \label{app:collab}}\begin{sloppypar}\hyphenpenalty=5000\widowpenalty=500\clubpenalty=5000
\cmsinstitute{Yerevan Physics Institute, Yerevan, Armenia}
{\tolerance=6000
A.~Hayrapetyan, A.~Tumasyan\cmsAuthorMark{1}\cmsorcid{0009-0000-0684-6742}
\par}
\cmsinstitute{Institut f\"{u}r Hochenergiephysik, Vienna, Austria}
{\tolerance=6000
W.~Adam\cmsorcid{0000-0001-9099-4341}, J.W.~Andrejkovic, T.~Bergauer\cmsorcid{0000-0002-5786-0293}, S.~Chatterjee\cmsorcid{0000-0003-2660-0349}, K.~Damanakis\cmsorcid{0000-0001-5389-2872}, M.~Dragicevic\cmsorcid{0000-0003-1967-6783}, P.S.~Hussain\cmsorcid{0000-0002-4825-5278}, M.~Jeitler\cmsAuthorMark{2}\cmsorcid{0000-0002-5141-9560}, N.~Krammer\cmsorcid{0000-0002-0548-0985}, A.~Li\cmsorcid{0000-0002-4547-116X}, D.~Liko\cmsorcid{0000-0002-3380-473X}, I.~Mikulec\cmsorcid{0000-0003-0385-2746}, J.~Schieck\cmsAuthorMark{2}\cmsorcid{0000-0002-1058-8093}, R.~Sch\"{o}fbeck\cmsorcid{0000-0002-2332-8784}, D.~Schwarz\cmsorcid{0000-0002-3821-7331}, M.~Sonawane\cmsorcid{0000-0003-0510-7010}, S.~Templ\cmsorcid{0000-0003-3137-5692}, W.~Waltenberger\cmsorcid{0000-0002-6215-7228}, C.-E.~Wulz\cmsAuthorMark{2}\cmsorcid{0000-0001-9226-5812}
% ...
\par}
\vskip\cmsinstskip
\dag:~Deceased\\
$^{1}$Also at Yerevan State University, Yerevan, Armenia\\
$^{2}$Also at TU Wien, Vienna, Austria\\
% ...
\end{sloppypar}
\end{document}